\documentclass[traditabstract]{aa}
\usepackage{graphicx}
\usepackage{txfonts}

\def\tex {\ifmmode{{T}_{\rm ex}}\else{$T_{\rm ex}$}\fi}
\def\tmb {\ifmmode{{T}_{\rm mb}}\else{$T_{\rm mb}$}\fi}
\def\ci     {\ifmmode{{\rm C}{\rm \small I}}\else{C\ts {\scriptsize I}}\fi}
\def\hi     {\ifmmode{{\rm H}{\rm \small I}}\else{H\ts {\scriptsize I}}\fi}
\def\hh     {\ifmmode{{\rm H}_2}\else{H$_2$}\fi}

\def\ts     {\thinspace}
\def\kms    {\ifmmode{{\rm \ts km\ts s}^{-1}}\else{\ts km\ts s$^{-1}$}\fi}
\def\msol   {\ifmmode{{\rm M}_{\odot}}\else{M$_{\odot}$}\fi}
\def\lsol   {\ifmmode{{\rm L}_{\odot}}\else{L$_{\odot}$}\fi}
\def\zsol   {\ifmmode{{\rm Z}_{\odot}}\else{Z$_{\odot}$}\fi}
\def\etal   {{\rm et\ts al.}~}

\begin{document}

\title{Molecular content of polar-ring galaxies
\thanks{Based on observations carried out with the IRAM 30m telescope.
IRAM is supported by INSU/CNRS (France), MPG (Germany), and IGN (Spain)}}

\author{F. Combes \inst{1}
\and
A. Moiseev \inst{2}
\and
V. Reshetnikov \inst{3}
           }
\offprints{F. Combes}
\institute{Observatoire de Paris, LERMA (CNRS:UMR8112), 61 Av. de l'Observatoire, F-75014, Paris, France
\email{francoise.combes@obspm.fr}
 \and
Special Astrophysical Observatory, Russian Academy of Sciences, 369167 Nizhnii Arkhyz, Karachaevo-Cherkesskaya Republic, Russia
 \and
St. Petersburg State University, Universitetskii pr. 28, 198504 St Petersburg, Stary Peterhof, Russia
              }

   \date{Received  2013/ Accepted  2013}

   \titlerunning{CO in Polar-Ring Galaxies}
   \authorrunning{F. Combes et al.}

   \abstract{We have searched for CO lines in a sample of 21 new morphologically
determined polar-ring galaxies (of which nine are kinematically confirmed),
obtained from a wide search in the Galaxy Zoo project by Moiseev and collaborators.
 Polar-ring galaxies (PRGs) are a unique class of objects, tracing special episodes in the galaxy mass
assembly: they can be formed through galaxy interaction and merging, but also through accretion
from cosmic filaments. Furthermore, they enable the study of dark matter haloes in three dimensions.
The polar ring itself is a sub-system rich in gas, where molecular gas is expected, and new stars are formed.
Among the sample of 21 PRGs, we have detected
five CO-rich systems, that can now be followed up with higher spatial resolution.
Their average molecular mass is 9.4 10$^9$ M$_\odot$, and their average gas fraction is
27\% of their baryonic mass, with a range from 15 to 43\%, 
implying that they have just accreted a large amount of gas.
 The position of the detected objects in the velocity-magnitude diagram 
is offset from the Tully-Fisher relation of normal spirals, as was already found for PRGs.
This work is part of our
multi-wavelength project to determine the detailed morphology and
dynamics of polar-ring galaxies, test through numerical models their formation
scenario, and deduce their dark matter content and 3D-shape.
}
\maketitle


\section{Introduction}

Polar-ring galaxies (PRGs, see Fig. \ref{fig:sample}) are peculiar objects composed of a
central component (usually an early-type galaxy) surrounded by
an outer ring or disk, made up of gas, stars, and dust, which
orbits nearly perpendicular to the plane of the gas-poor central
galaxy (Whitmore \etal 1990).
 For some well-known systems studied with high spatial resolution, it
was possible to show that the polar system is in fact a polar disk, more than 
a polar ring (e.g. Iodice \etal 2002b). However, polar disks and rings are 
both gathered into the same PRG class.

Measurements of their kinematics can therefore give some insight
in the 3D-shape of their dark matter, which can be generalised to
the progenitor spiral galaxies, provided that their formation mechanism is known.

We note that many of the best cases of PRGs have a relatively massive polar component
that cannot be treated as simple test particles, but self-gravity must be taken into account.

\subsection{Formation scenarios for polar-ring galaxies}

From dynamical arguments, the two misaligned systems cannot be
formed simultaneously, as confirmed by the younger ages of the
polar rings/disks (Iodice \etal 2002a, b), and the fact that most
of the gas of the system is in the polar disk (van Driel \etal
2000, 2002). At least three formation mechanisms have been
discussed in the literature, the first two involving galaxy interactions:

1) the accretion scenario, where two interacting galaxies exchange
mass, as invoked by Schweizer
\etal (1983) and simulated by Reshetnikov \& Sotnikova (1997);

2) the merging scenario, or the head-on collision of two orthogonal
spiral galaxies, first studied by Bekki (1997, 1998).
Bournaud \& Combes (2003) have shown through simulations
that statistically the first scenario is more frequent, and more likely
to represent observations;

3)  the cosmic formation scenario, where the PRGs form through the misaligned accretion
of gas from cosmic filaments (Maccio \etal 2006; 
Brook \etal 2008). The gas of the PR is then of lower metallicity 
than in the first scenarios.

 The fact that some PRGs are polar disks more than polar rings
supports the formation through cosmic accretion, as shown by Spavone \etal (2010).
Also, the low metallicity, and flat abundance gradients could favour
this mechanism (Spavone \etal 2011), while the presence of a true ring supports
the tidal accretion.

Snaith \etal (2012) have developed in more detail the formation scenario proposed by 
Brook \etal (2008). In their simulations, the polar disk is progressively
formed out of gas and dark matter infalling from a cosmic filament, after 
the last major merger has re-oriented the old system in a perpendicular
direction (and formed the host). This implies that the dark matter is aligned
with the polar system.

\subsection{The dark matter issue}

 The 3D-shape of dark matter halos has been estimated in 
many objects (e.g. Combes 2002), and has recently been investigated
through edge-on galaxies, allowing us to study the gas flaring of the disk
(e.g. O'Brien \etal 2010, Peters \etal 2013), through gravitational lensing
(van Uitert \etal 2012), or through satellite disruption (Law \etal 2009).
There is a large scatter in the flattening derived, i.e.
an axis ratio between 0.2 and 1.
Polar-ring galaxies are privileged systems for this study.
Since the ring is gas rich and rotates around the pole, it is
a probe of the gravitational potential in the third
dimension, which is normally inaccessible in normal spirals.
It is therefore a unique tool for determining the 3D-shape
and consequently constraining the nature of dark matter.
From the shape of dark haloes in PRGs, we should be able to deduce
the shape of dark haloes in normal galaxies,  knowing
the formation mechanisms of PRGs.

In all previous studies (e.g. Whitmore \etal 1987, Sackett \& Sparke 1990,
Sackett \etal 1994, Reshetnikov \& Combes 1994, Combes \& Arnaboldi 1996,
Iodice \etal 2003) the common conclusion is that PRGs are
indeed embedded in a dark halo.
However, the solutions for the 3D-shape
differ, according to models and accuracy of data:
the dark halo is almost spherical for Whitmore \etal (1987), flattened
along the equatorial plane of the host galaxy (Sackett \etal 1994),
or flattened along the polar-ring plane (Combes \& Arnaboldi 1996).
 This last solution is supported for a large number of PRGs
by Iodice \etal (2003), through a study of the Tully-Fisher (TF) diagram
for PRGs.

The position of the PRGs in the TF diagram is very peculiar,
and does not fit the sequence of normal disk galaxies (Iodice \etal 2003, Reshetnikov 2004). The rotational
velocity is obtained through HI-21cm measurement from the gas in the
edge-on polar disks.
Most of the PRGs have higher than normal rotational velocities,
for their luminosities.  This is not expected  if the dark halo
is spherical or flattened to the equatorial plane of the host,
because then the observed velocity corresponds
to the apocenter of the excentric polar orbit, and is lower
than the velocity observed in the equatorial plane.

Since the contrary is observed, this must be due to a flattening
of the dark matter towards the polar plane.  This
important suggestion must be confirmed by a much larger statistics,
and therefore we want to enlarge significantly the number of
objects that can be considered as PRGs. Since this implies selecting
objects that are more distant because the HI-21cm line is not as easy to detect
at high redshift, the CO line then becomes the preferred tracer of the gas.
In the present sample, our largest redshift is 0.078, but in the future with
ALMA, the CO-TF will be a unique tool.

\subsection{Molecular gas in PRGs}

 The molecular content of PRGs is poorly known. 
The first CO detection in such an object was in NGC 660
(Combes \etal 1992). Watson \etal (1994) then
detected CO(2-1) in the polar rings of NGC 2685 and NGC 4650A, and later
Schinnerer \& Scoville (2002) made an interferometric map of the Spindle galaxy
NGC 2685. Van Driel \etal (1995) found abundant CO emission in the
inclined ring of NGC 660,
and Crocker \etal (2008) in the center of NGC 2768.
Galletta \etal (1997) observed ten PRGs in CO and found molecular masses
much larger than those in early-type galaxies and also
in dwarf galaxies, suggesting that PRGs cannot get their gas in
 a dwarf accretion only. 
 All these are only a few cases, and more observations are required
 to know better the molecular content of polar-ring
galaxies and to better understand their formation scenarios.

The sample is described in Sect. \ref{sample} and the observations
in  Sect. \ref{obs}. Results are presented in  Sect. \ref{res} and
discussed in Sect. \ref{disc}. 

\begin{figure*}[Hp]
\centerline{
\includegraphics[angle=-0,width=16cm]{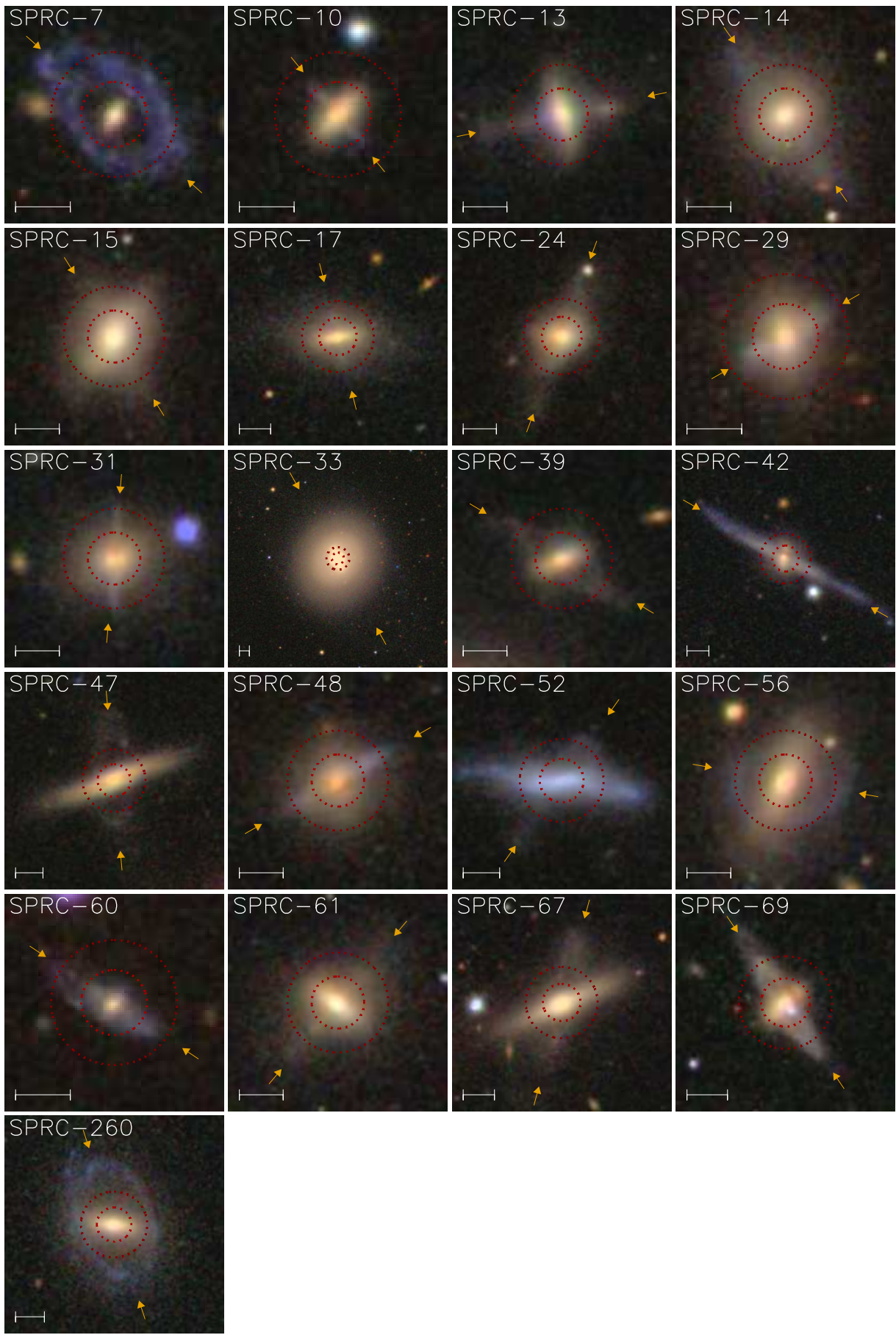}}
\caption{SDSS DR8 colour images of the polar ring sample (see Table \ref{tab:sample}). 
The scale bar in each image is 10 arcsec long, and the two dashed circles indicate
the Full Width at Half Power (FWHP) beam width for the CO(1-0) and CO(2-1) observations.  To identify the polar systems,
we have indicated their directions with arrows, corresponding to the PA
displayed in Table \ref{tab:sample}. SPRC-33
 contains a giant HI polar disk with very weak stellar counterpart detected
 in UV images only (Bettoni \etal 2010, Moiseev \etal 2011).
}
\label{fig:sample}
\end{figure*}

\section{The sample}
\label{sample}

Recently, Moiseev \etal (2011) have built a new catalogue of PRGs,
significantly increasing the number of known candidate PRGs. The
catalogue is based on the results of the original Galaxy Zoo project,
where nearly a million galaxies from the Sloan Survey (SDSS) were classified.
This results in the Polar Ring Catalog (SPRC) of 275 objects,
in which 70 galaxies are classified as the best
 PRGs, and 115 as good PRG candidates.  Among the 70,
there are 15 kinematically confirmed PRGs. For our search, we
have selected the latter, and among the best candidates, the brightest
in terms of r-magnitude (r$<$ 15.5) (cf. Table \ref{tab:sample}).

The main goal of the present work is to obtain information
on the molecular content of this sample of 21 PRGs.
 The global CO detection allows us to locate the object in the Tully-Fisher
diagram, and to have a first information in the geometry of the dark halo.
 Future interferometric work on the detected PRGs will provide high resolution maps, in order
to be able to compare results with numerical models.

\begin{figure}[h]
\resizebox{8cm}{!}{\includegraphics[angle=0,width=8cm]{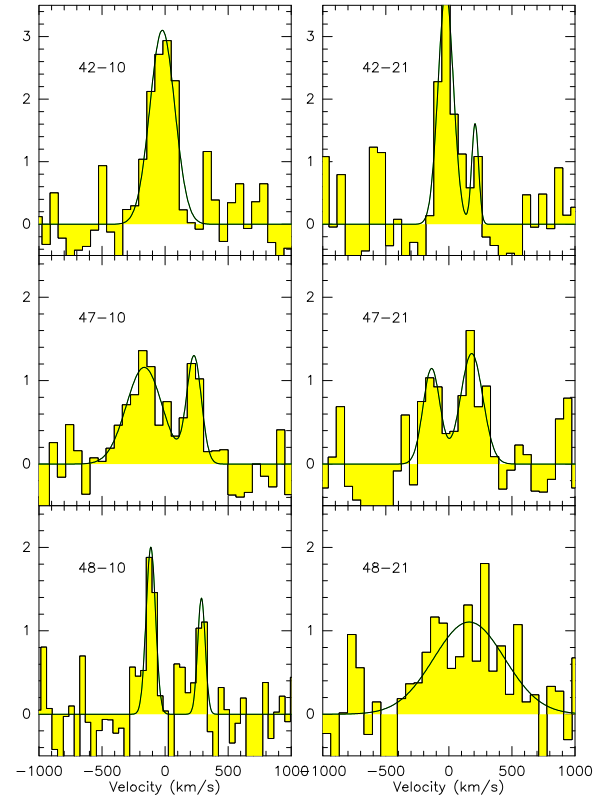}}
\caption{CO spectra of the detected PRG galaxies, not yet kinematically confirmed. 
Left-hand panels are the CO(1-0), right-hand panels CO(2-1) emission. The
velocity scale is relative to the redshift displayed in Table \ref{tab:sample}.
The vertical scale is T$_{mb}$ in mK.}
\label{fig:spec1}
\end{figure}

\begin{figure}[h]
\resizebox{8cm}{!}{\includegraphics[angle=0,width=8cm]{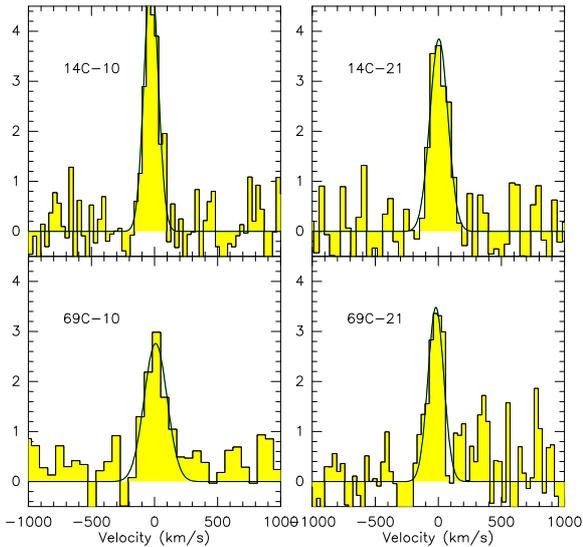}}
\caption{Same as Fig. \ref{fig:spec1}  for the two kinematically confirmed and CO-detected PRGs. }
\label{fig:spec2}
\end{figure}

In this article, we adopt a standard flat cosmological model,
with $\Lambda$ = 0.73 and a Hubble constant of 71\,km\,s$^{-1}$\,Mpc$^{-1}$
(Hinshaw \etal  2009).

%
\begin{center}
\begin{table*}
      \caption[]{Basic data for the polar-ring galaxy sample, 
and sizes of the regions observed in CO(1-0) and CO(2-1)}
\label{tab:sample}
\begin{tabular}{llrrrrrrrrr}
\hline
SPRC &  Name  & R.A.  &  Dec.  & $cz$ & g& r & i&PA &beam$_{10}$ & beam$_{21}$\\
 No.$^{(1)}$ &  & \multicolumn{2}{c}{(2000.0)} & [km/s]&[mag]&[mag]&[mag]&[$^\circ$]&[kpc]&[kpc]\\
\hline
7C &            &07 52 34.32 &  +29 20 49.7 &  18032 & 17.84& 16.95 &16.85&48 &27.9 &14.0\\
10C&            &08 20 38.19 &  +15 36 59.8 &  12736 & 17.16& 16.29 &15.79&39 &19.8 & 9.9\\
13 &            &09 14 53.66 &  +49 38 24.0 &  9521  & 16.10& 15.37 &15.06&102&14.8 & 7.4\\
14C&CGCG 121-053&09 18 15.97 &  +20 22 05.3 &  9548  & 15.59& 14.73 &14.48&35 &14.8 & 7.4\\
15 &            &09 36 34.63 &  +21 13 57.8 & 10281  & 15.31& 14.43 &14.14&32 &16.0 & 8.0\\
17 &            &09 59 11.85 &  +16 28 41.5 &  7914  & 15.81& 14.92 &14.59&14 &12.3 & 6.2\\
24 &            &11 16 25.11 &  +56 50 17.0 & 14133  & 15.85& 14.98 &14.63&160&21.9 & 11.0\\
29 &            &11 53 33.56 &  +47 19 07.3 & 14208  & 16.30& 15.25 &15.06&120&22.1 & 11.0\\
31 &            &12 17 11.51 &  +31 30 37.8 & 14913  & 16.21& 15.08 &14.93&175&23.1 & 11.6\\
33C&  NGC 4262  &12 19 30.57 &  +14 52 39.5 & 1358   & 12.22& 11.24 &10.98&29 &21.1 & 10.6\\
39C&            &13 08 16.92 &  +45 22 35.2 & 8792   & 17.01& 16.01 &15.75&59 &13.7 & 6.8\\
42 &  UGC 08634 &13 39 04.59 &  +02 09 49.5 &  7041  & 15.75& 15.04 &14.85&60 &11.0 & 5.5\\
47 &            &13 59 41.70 &  +25 00 46.1 &  9370  & 15.62& 14.49 &14.28&5  &14.6 & 7.3\\
48 &            &14 14 20.82 &  +27 28 04.4 & 16788  & 16.30& 15.04 &15.10&120&26.0 & 13.0\\
52 &KUG 1416+257&14 18 25.60 &  +25 30 06.7 &  4450  & 15.70& 15.07 &14.50&145& 6.9 & 3.5\\
56 &MCG +06-33-026&15 11 14.09& +37 02 37.7 & 16499  & 15.69& 14.81 &14.49&80 &25.6 & 12.8\\
60C&            &15 47 24.32 &  +38 55 50.4 & 23519  & 17.71& 17.29 &17.02&56 &36.3 & 18.1\\
61 &            &15 49 54.81 &  +09 49 43.1 & 13753  & 15.78& 14.81 &14.50&140&21.4 & 10.7\\
67C&CGCG 225-097&17 17 44.13 &  +40 41 52.0 &  8325  & 15.28& 14.25 &14.01&165&12.9 & 6.5\\
69C& II Zw 092  &20 48 05.67 &  +00 04 07.8 &  7396  & 16.22& 15.36 &14.93&34 &11.6 & 5.8\\
260C&CGCG 068-056&11 45 30.25&  +09 43 44.8 &  6399  & 15.31& 14.48 &14.23&18 &10.0 & 5.0\\
\hline
\end{tabular}
\\$^{(1)}$The numbers are followed by C when kinematically confirmed.
\\ PA is the position angle of the polar ring (with uncertainty $\pm$5$^\circ$).
\\The kinematics were obtained for SPRC-7 by Brosch \etal (2010),
for SPRC-33 by Bettoni \etal (2010),
for SPRC-67 by Merkulova  \etal (2012),
for SPRC-10, 14, 39, 60, and 69 by Moiseev \etal (2011),
for SPRC-260, by Khoperskov \etal (2012).
\\ Magnitudes are from SDSS.
\end{table*}
\end{center}

\section{Observations}
\label{obs}

The observations were carried out with the IRAM 30m telescope
at Pico Veleta, Spain, from December 2011 to January 2012.
All sources were observed simultaneously in CO(1-0) and
CO(2-1) lines, with the 3mm and 1mm receivers in parallel.

The broadband EMIR receivers were tuned in single sideband mode,
with a total bandwidth of 4 GHz per polarization. This covers a velocity range
of $\sim$ 10,400 \kms at 2.6mm and $\sim$ 5,200 \kms at 1.3mm.  The
observations were carried out in wobbler switching mode, with reference
positions offset by $2\arcmin$ in azimuth. Several backends were used in parallel,
the WILMA autocorrelator with  $2$~MHz channel width, covering 4$\times$4\,GHz,
and the $4$~MHz filterbanks, covering 2$\times$4\,GHz.

We spent on average two hours on each galaxy, and reached
a noise level between 0.6 and 1.6 mK (antenna temperature), smoothed
over  $30$~km~s$^{-1}$ channels for all sources. 
Pointing measurements were carried out every two hours on continuum sources and
the derived pointing accuracy was 3$''$ rms.  The temperature scale is then
transformed from antenna temperature  $T_{\rm A}^*$ to main
beam temperature $T_{\rm mb}$, by multiplying by 1.17 at 3mm and 1.46 at 1.3mm. 
To convert the signals to fluxes, we use $S/T_{\rm mb}$ = 5.0 Jy/K for all bands. 
At 2.6mm and 1mm, the telescope half-power beam
width is 23$''$ and 12$''$ respectively. 
The data were reduced with the CLASS/GILDAS software, and
the spectra were smoothed so that each line covers about ten channels in the plots.
\footnote{Spectra of detections are available in electronic form
at the CDS via anonymous ftp to cdsarc.u-strasbg.fr (130.79.128.5)
or via http://cdsweb.u-strasbg.fr/cgi-bin/qcat?J/A+A/
}

\begin{table}
      \caption[]{Molecular data and stellar mass of the five detected PRGs.
For each object, the first line displays the CO(1-0) and the second line the CO(2-1) results.}
\label{tab:det}
\begin{center}
\begin{tabular}{lcccccc}
\hline
%
%
\scriptsize{SPRC}& Area &  V & $\Delta$V$^{(1)}$ & T$_{mb}^{(2)}$ & M(H$_2$)$^{(3)}$& M*\\
  No.  &  K km/s &   km/s &   km/s  & mK &  \multicolumn{2}{c}{10$^9$ M$_\odot$}\\
\hline
14C&  0.81$\pm$ 0.1  & 5$\pm$ 11 &165$\pm$ 24 &4.6 & 8.4 & 23\\
    &  1.04$\pm$ 0.1  & -26$\pm$ 5  &127$\pm$ 12 &7.7 && \\
69C&  0.71$\pm$ 0.1  &6$\pm$ 15  &193$\pm$ 40 &3.5 & 4.5& 11\\
    &  0.68$\pm$ 0.1  &-17$\pm$ 10  &128$\pm$ 23 &5.0 && \\
42 &  0.91$\pm$ 0.1  &-22$\pm$ 16  &231$\pm$ 34 &3.6 & 5.2& 7 \\
    &  0.82$\pm$ 0.2  &-24$\pm$ 14  &131$\pm$ 36 &5.8 && \\
47 &  0.73$\pm$ 0.1  &-59$\pm$ 61  &550$\pm$ 110&1.3 & 7.3& 40\\
    &  0.72$\pm$ 0.2  & 74$\pm$ 57  &479$\pm$ 95 &1.5 && \\
48 &  0.65$\pm$ .05  & 21$\pm$ 17  &380$\pm$ 50 &1.7 & 21.5& 70 \\
    &  0.72$\pm$ 0.2  &142$\pm$ 67  &450$\pm$ 143&1.6 && \\
\hline
\end{tabular}
\end{center}
\begin{list}{}{}
\item[]   Results of the Gaussian fits
\item[] $^{(1)}$ FWHM 
\item[] $^{(2)}$ Peak brightness temperature
\item[] $^{(3)}$ obtained with the standard MW conversion ratio
\end{list}
\end{table}

\begin{figure}[h!]
\centerline{
\includegraphics[width=8cm]{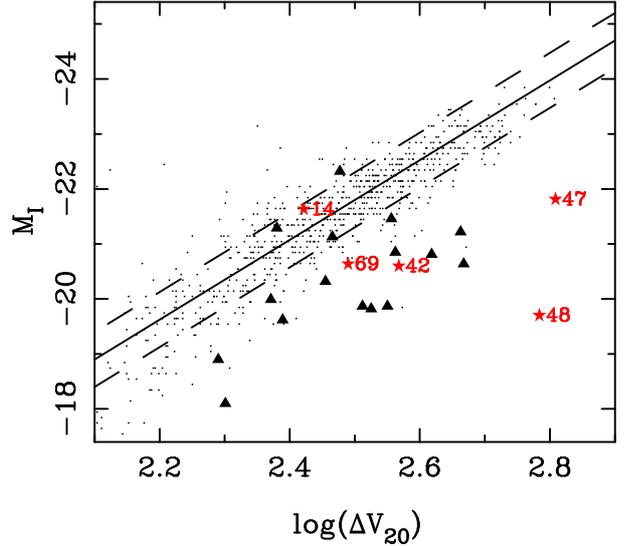}
}
\caption{Tully-Fisher relation for our detected PRGs (red stars), compared with other
PRGs (black triangles, Whitmore \etal 1990, van Driel \etal 2002, 
Iodice \etal 2003), and normal spirals (black dots, Giovanelli \etal 1997).
The I absolute magnitudes, with uniform photometry and distance corrections,
are plotted vs. observed rotation velocities (full width at 20\% power), in the polar rings. 
}
\label{fig:TF}
\end{figure}

\begin{table}
      \caption[]{Upper limits}
         \label{tab:uplim}
            \begin{tabular}{l c c c c c }
            \hline
            \noalign{\smallskip}
SPRC &Line&$\nu_{\rm obs}$& rms & L'$_{\rm CO}$/10$^{9}$& M(H$_2$) \\
 No.    &     & [GHz]     &  [mK]  & [K \kms\, pc$^2$] &  [10$^9$\msol]\\
            \noalign{\smallskip}
            \hline
            \noalign{\smallskip}
   7C  &  CO(10)  & 108.7  &  0.8  &  1.16  & 5.3 \\
        &  CO(21)  & 217.4  &  1.6  &  0.72  & 3.3 \\
   10C &  CO(10)  & 110.6  &  0.6  &  0.43  & 2.0 \\
        &  CO(21)  & 221.1  &  1.1  &  0.25  & 1.1 \\
   13  &  CO(10)  & 111.7  &  0.9  &  0.36  & 1.6 \\
        &  CO(21)  & 223.4  &  1.2  &  0.15  & 0.7 \\
   15  &  CO(10)  & 111.4  &  0.7  &  0.33  & 1.5 \\
        &  CO(21)  & 222.9  &  0.8  &  0.12  & 0.5 \\
   17  &  CO(10)  & 112.3  &  0.7  &  0.19  & 0.9 \\
        &  CO(21)  & 224.6  &  0.7  &  0.06  & 0.3 \\
   24  &  CO(10)  & 110.1  &  0.8  &  0.71  & 3.2 \\
        &  CO(21)  & 220.1  &  1.4  &  0.38  & 1.8 \\
   29  &  CO(10)  & 110.0  &  0.7  &  0.63  & 2.9 \\
        &  CO(21)  & 220.1  &  1.1  &  0.31  & 1.4 \\
   31  &  CO(10)  & 109.8  &  0.7  &  0.69  & 3.2 \\
        &  CO(21)  & 219.6  &  1.5  &  0.46  & 2.1 \\
   33C &  CO(10)  & 114.7  &  1.1  &  0.90  & 4.1 \\
        &  CO(21)  & 229.5  &  0.9  &  0.23  & 1.0 \\
   39C &  CO(10)  & 111.9  &  0.7  &  0.24  & 1.1 \\
        &  CO(21)  & 223.9  &  0.7  &  0.07  & 0.3 \\
   52  &  CO(10)  & 113.6  &  0.9  &  0.08  & 0.3 \\
        &  CO(21)  & 227.2  &  1.1  &  0.03  & 0.1 \\
   56  &  CO(10)  & 109.2  &  0.9  &  1.09  & 5.0 \\
        &  CO(21)  & 218.5  &  1.3  &  0.49  & 2.2 \\
   60C &  CO(10)  & 106.9  &  0.6  &  1.49  & 6.9 \\
        &  CO(21)  & 213.8  &  0.9  &  0.70  & 3.2 \\
   61  &  CO(10)  & 110.2  &  0.6  &  0.50  & 2.3 \\
        &  CO(21)  & 220.4  &  1.1  &  0.29  & 1.3 \\
   67C &  CO(10)  & 112.1  &  0.7  &  0.21  & 1.0 \\
        &  CO(21)  & 224.3  &  1.0  &  0.09  & 0.4 \\
  260C &  CO(10)  & 112.9  &  1.0  &  0.18  & 0.8 \\
        &  CO(21)  & 225.7  &  0.9  &  0.05  & 0.2 \\
            \noalign{\smallskip}
            \hline
           \end{tabular}
\begin{list}{}{}
\item[]  The rms are in T$_{\rm A}^*$ in  channels of 30 \kms.
\item[] The upper limits in L'$_{\rm CO}$ and M(\hh) are at 3$\sigma$ with an assumed $\Delta$V = 300 \kms. 
\end{list}
\end{table}

\section{Results}
\label{res}

\subsection{CO detection in PRGs}
\label{COdet}

Figures \ref{fig:spec1} and \ref{fig:spec2} 
display the CO-detected sources, in both CO lines. Two of the five
detections involved kinematically confirmed PRGs.
 Table \ref{tab:det} reports
all line parameters for the detections, and the upper limits for the non-detections are
reported in Table  \ref{tab:uplim}.
Integrated signals and velocity widths have been computed from Gaussian fits.
These also give the central velocities, with respect to the optical
redshift of Table \ref{tab:sample}. 
The upper limits are computed at 3$\sigma$,
assuming a common line width of 300\kms and getting
the rms of the signal over 300\kms. 

The detection rate of 24$\pm$9\% is comparable to that found for early-type galaxies,
for which the gas content is believed to be due to recent accretion,
as was found in the SAURON and ATLAS$^{\rm 3D}$ samples (e.g. Combes \etal 2007, 
Young \etal 2011).

One problem in the interpretations of the data is the size of the CO beams, 
which are sometimes smaller than the total extent of the polar rings.
Table \ref{tab:sample} lists the values of the two beams in kpc
on the major axis of the galaxies.
For the CO(1-0) beam, two detected galaxies, SPRC-42 and SPRC-47, 
are particularly in this case,
where the H$_2$ mass derived in Table \ref{tab:det} might 
only be a lower limit. The shape of the CO(1-0) and CO(2-1) velocity profiles are not the
same, since both lines resolve the polar ring differently.
Among the upper limits, four polar rings are also in this case,
SPRC-33, 52, 67 and 69. A full map would be required to settle the issue. 

In all the detected galaxies, we interpret the CO emission as coming from
the polar-ring system, since it is always the bluer one (cf. Fig. \ref{fig:sample}).
 All five of the detected polar rings are almost edge-on, 
which minimizes the errors done in correcting for inclination
the observed linewidths, to obtain indicative 
rotational velocities (cf. Fig. \ref{fig:TF}).

  The line widths detected are in average 304\kms\, Full Width at Half Maximum (FWHM). Two of the PRGs, P47 and P48,
 clearly show double-horn profiles, indicative of rotating disks or rings. 
For two of the detected galaxies, the rotational velocity has also been estimated  
in the ionized gas (Moiseev \etal 2011). 
For SPRC-14 the ionized gas data in projection on the sky has a
symmetrized maximal line-of-sight rotation velocity of 
160 km/s, in good
agreement with the CO(1-0) value of 165$\pm$24 km/s.
The polar ring is, however, asymmetric, and with an extension on the NE side,
where the velocity reaches up to 
Vmax= 262$\pm$11 km/s (H$\alpha$) and  268$\pm$5 (NII).  

For SPRC-69 the ionized gas velocity field yields
Vmax=178$\pm$3 km/s, (172 after projecting with the inclination 
of 76$^\circ$ of the polar ring)
a value also in agreement with the CO(1-0) value 193$\pm$40 km/s.

These comparisons show that we are not underestimating too much
the maximum velocities of the polar rings, even though the CO beam
might not encompass all the optical extent of it. This might come
from the fact that the maximum velocity is reached already at small radii, 
especially for the 
molecular gas, which is more concentrated towards the center.
 We have plotted our detected objects in the Tully-Fisher diagram
of Fig. \ref{fig:TF}, in comparison with normal spirals
from Giovanelli \etal (1997) and other PRGs from Iodice \etal (2003). We note
that we have corrected the velocity for AM 2020-504, which is erroneous in
Iodice \etal, replacing it with the value from Whitmore \etal (1990) and van Driel \etal (2002).
We have converted the $i$ to $I$ magnitudes by the formula
\begin{eqnarray}\nonumber
I-i &=& -0.0307(r-i)^3 +0.1163(r-i)^2 \\ & &
\nonumber
-0.3341(r-i)-0.3584
\end{eqnarray}
from Ivezic \etal (2007).
 The CO-detected PRGs show the same tendency to lie to the right
of the main relation, i.e. their velocities are too high. Since the velocities
determined from CO are lower limits (because of the restricted beam), this result
is robust.

\subsection{CO luminosity and \hh\, mass}
\label{CO-mass}

We have simultaneously observed the two first lines of the CO
rotational ladder, and it is interesting to compare them,
to have an idea of the excitation of the gas.
Therefore, we compute L'$_{CO}$, the special unit CO luminosity,
through integrating the CO intensity over the velocity profile.
  This luminosity, expressed in units of K \kms pc$^2$,
 will give the same value irrespective of  J, if the
CO lines are saturated and have the same brightness temperature.

This CO luminosity is given by $$L'_{CO} =
23.5 I_{CO} \Omega_B {{D_L^2}\over {(1+z)^3}} \hskip6pt \rm{K\hskip3pt
  km \hskip3pt s^{-1}\hskip3pt pc^2}$$,
where  $I_{CO}$ is the intensity in K \kms, $\Omega_B$ the area of
the main beam in square arcseconds, and $D_L$ the luminosity
distance in Mpc.  

In most cases (15 out of the 21 objects in the sample) 
the beam in CO(1-0) is large enough that it is likely to encompass
all the emission of the polar-ring system; however, it is not the same
in CO(2-1), and only the CO(1-0)-derived H$_2$ masses should
be trusted. 
For the detected objects, the integrated intensities are always
comparable between the CO(1-0) and CO(2-1) lines, as can be seen in 
Table \ref{tab:det}. For point sources, with saturated and thermalized CO lines,
the ratio should be as high as 4, in favour of the CO(2-1). The 
fact that intensities are comparable can be interpreted either
in terms of an extended emission, or a sub-thermal excitation, or both.

We have computed the molecular mass from the CO(1-0) flux,
 using
M$_{\rm H_2} = \alpha$ L'$_{\rm CO}$, 
with $\alpha=4.6$ M$_\odot$ (K \kms\, pc$^2$)$^{-1}$,
the standard factor for nearby quiescent galaxies like the Milky Way. 
The molecular gas masses are
listed in Table~\ref{tab:det} and the upper-limits in  Table~\ref{tab:uplim}. 

The average CO luminosity for the five galaxies detected
is L'$_{\rm CO}$ = 2.0 10$^{9}$ K \kms\, pc$^2$, corresponding to an 
average \hh\, mass of  9.4 10$^{9}$ \msol.
These gas masses are relatively high, and we now compare
them to stellar masses for each system.

\subsection{Stellar mass and gas fraction}
\label{GFR}

We compute stellar masses from observed optical (SDSS) and near infrared
(2MASS) magnitudes, 
using calibrated relations between mass-to-light ratios and colours
(see e.g. Bell \etal 2003).  The multi-wavelength luminosities
were K-corrected according to the colours (cf. Chilingarian \etal 2010). 
Stellar masses are displayed in Table \ref{tab:det}.
The gas fractions derived from these
stellar masses show large variations, 
between 15\% and 43\%, with an average of 27\%. These gas fractions
are quite high, compared to normal spirals at the same redshifts.
The selection of bright polar-ring galaxies at these distances
means therefore the selection of galaxies that have just accreted a large amount of
gas mass.

\section{Summary and discussion}
\label{disc}

We have presented
 our CO survey in 21 polar-ring galaxies, observed with the IRAM-30m telescope. Five
galaxies were detected, and among them two kinematically confirmed PRGs.
The detection rate of 24$\pm$9\% is comparable to early-type galaxies, where the gas
is thought to have been recently accreted.
The first two CO lines were observed, and the L'(CO) luminosities in CO(2-1) are
lower than in CO(1-0), indicating either a subthermal gas, and/or a gas extent
larger than the $\sim$ 12'' CO(2-1) beam.
 Assuming a standard CO-to-\hh\, conversion factor, the average molecular gas
mass is found to be  9.4 10$^{9}$ \msol.
 The average ratio between gas and stellar mass is 0.4, or the average gas fraction
is 27\%. This high fraction means that bright polar-ring galaxies have just accreted
a large amount of gas.

 We interpret our CO detections as coming from the polar-ring systems in 
the detected galaxies, since it is the bluer and younger component.
Deriving the rotational velocity of the gas from the CO profile, we
have placed our observed galaxies in the Tully-Fisher diagram, allowing us
to compare them with the control sample of normal spirals.
 The new detected objects confirm the offset position of PRGs already
noticed by Iodice \etal\, (2003).

 In Fig. \ref{fig:TF}, one of the PRGs falls in the expected range
for a spherical potential (SPRC-14), two are mildly offset (SPRC-69 and SPRC-42),
and two are very far from the usual relation, with very broad velocities
(SPRC-48 and SPRC-47). This large scatter might indicate different
formation mechanisms for the systems, and at least four out of the five should have
their dark matter halo aligned with the polar plane. 
The position of the five PRGs in the TF diagram does not appear related to 
the mass ratio between the polar disk/ring and the host. The galaxies SPRC-42 and  SPRC-69 have 
polar structures brighter than their hosts, while  SPRC-14 and  SPRC-48 are comparable,
and SPRC-47 is weaker than its host. This was also the case in the previous
PRGs (Iodice \etal 2003). There is no obvious relation with the PR spatial extent either.

The two most extreme cases
(SPRC-47 and SPRC-48) must have a very flattened dark-matter system,
as flat as a disk. This result is however uncertain, since the mass distribution
is not yet known, and the CO beam does not cover the whole polar disk: the velocity
could be higher in the center, and the flat portion of the rotation curve be lower.
A full CO map at high spatial resolution is required to conclude.

\begin{acknowledgements}
 We warmly thank the referee for constructive comments and suggestions. 
The IRAM staff is gratefully acknowledged for their
help in the data acquisition. 
F.C. acknowledges the European Research Council
for the Advanced Grant Program Number 267399-Momentum.
A.M. is also grateful to the ‘Dynasty’ Foundation and RFBR grant 13-02-00416.
V.R. acknowledges partial financial support from the RFBR grant
11-02-00471−a.
We made use of the NASA/IPAC Extragalactic Database (NED),
and of the HyperLeda database.
\end{acknowledgements}

\end{document}